\documentclass{rnaastex}

\usepackage{CJK}

\begin{document}

\begin{CJK}{UTF8}{gbsn} 

\title{Super-Flaring Active Region 12673 Has One of the Fastest Magnetic Flux Emergence Ever Observed}

\correspondingauthor{Xudong Sun}
\email{xudongs@hawaii.edu}

\author[0000-0003-4043-616X]{Xudong Sun (孙旭东)}
\affiliation{Institute for Astronomy, University of Hawaii at Manoa, Pukalani, HI 96768-8288, USA}

\author[0000-0003-2622-7310]{Aimee A. Norton}
\affiliation{W. W. Hansen Experimental Physics Laboratory, Stanford University, Stanford, CA 94305-4085, USA}


\section{} 

In September 2017, solar active region (AR) 12673 evolved from a decaying sunspot to the most actively flaring region of Cycle 24. It produced 4 flares above \textit{GOES} X1 class and 8 flares above M3 class. The total flare index surpasses that of the great AR 12192 despite having only $\sim$30\% the sunspot size \citep{sun2015}. Its X-9.3 (\texttt{SOL2017-09-06T11:53}) flare is the most intense one since 2005.

\begin{figure}[th!]
\begin{center}
\includegraphics[scale=1.0,angle=0]{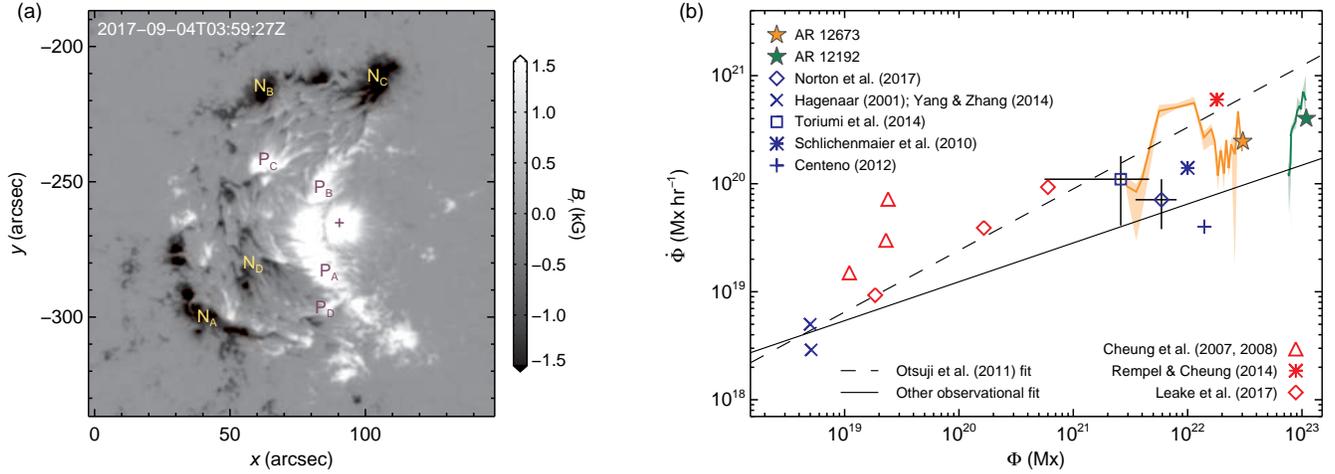}
\caption{Rapid flux emergence in AR 12673. (a) HMI radial field map. Following \citet{yangsh2017}, we mark the old sunspot by a plus sign, and four major bipoles, i.e. positive (negative) component of bipole A by $\mathrm{P_A}$ ($\mathrm{N_A}$), etc. $\mathrm{P_A}$ and $\mathrm{P_B}$ have coalesced into an arc, hugging the sunspot from the east. The collision site between the newly emerging $\mathrm{P_C}$ and $\mathrm{N_D}$ is the future epicenter of flaring activity. (b) Diagram of mean signed $\dot\Phi$ vs maximum signed $\Phi$ for various observations and simulations, adapted from Figure~5 of \citet{norton2017}. Red and blue symbols indicate simulations and observations, respectively. We overplot AR 12673 and its evolutionary track in orange, and AR 12192 in green. We simply divide the unsigned $\Phi$ and $\dot\Phi$ by 2 to get the signed version.\label{fig1}}
\end{center}
\end{figure}

The flaring activity is related to the emergence of significant new magnetic flux. Figure~\ref{fig1}(a) shows a radial field map from the Helioseismic and Magnetic Imager \citep[HMI;][]{schou2012}. The emergence started southeast of an old sunspot (remnant from AR 12665 and 12670). Multiple pairs of bipoles subsequently emerged on the eastern periphery of the old sunspot with very different orientations. Persistent flux emergence, apparent coalescence, cancellation, and shearing led to a complex $\beta\gamma\delta$ sunspot group, featuring a quadrupolar configuration and a reversed sigmoidal main polarity inversion line. A detailed account of the morphological evolution is available in \citet{yangsh2017}.
 
Using HMI vector magnetograms \citep{hoeksema2014} from September 2$^\mathrm{nd}$ to 7$^\mathrm{th}$, we calculated\footnote{We use pixels where the field strength is above $200~\mathrm{G}$. The uncertainty of $\Phi$ shows the standard deviation within a 96-min averaging window. The formal uncertainty from spectral inversion is small. We calculate $\dot\Phi$ using linear regression on bootstrapped samples of $\Phi$, and show the median and $2\sigma$ confidence interval.} the unsigned flux ($\Phi$) and the flux emergence rate ($\dot\Phi$). We find\footnote{Data quality deteriorates after September 7$^\mathrm{th}$ when the AR is more than 45$^\circ$ away from the disk center. We take $\Phi$ at 2017-09-07T00:00 as the maximum flux, $\Phi_\mathrm{m}\approx6.08\times10^{22}~\mathrm{Mx}$, and the mean before 2017-09-02T12:00 as the background flux $\Phi_0\approx6.20\times10^{21}~\mathrm{Mx}$. Following \citet{leka2013}, we define 2017-09-03T11:00 as the start of emergence when $\Phi-\Phi_0=0.1(\Phi_\mathrm{m}-\Phi_0$).} that $\Phi$ has a tenfold increase over the 5-day period to reach $6.08^{+0.02}_{-0.02}\times10^{22}~\mathrm{Mx}$, and the average $\dot\Phi$ is $4.93^{+0.11}_{-0.13}\times10^{20}\mathrm{~Mx~hr^{-1}}$. We additionally calculated $\dot\Phi$ in 6-hr chunks. The ``instantaneous'' flux emergence rate peaked around September 3$^\mathrm{rd}$ 21:00 with an extraordinary value of $1.12^{+0.15}_{-0.05}\times10^{21}~\mathrm{Mx~hr^{-1}}$. This occurred early during the evolution, when the flux is only 36\% of the maximum.

The flux emergence rate of AR 12673 is greater than any values reported in the literature of which we are aware. Figure~\ref{fig1}(b) summarizes the mean signed flux emergence rate and peak signed flux from a collection of observations and simulations \citep{norton2017}. AR 12763 largely follows the established trend, but the extreme values on its evolutionary track are rivaled only by the fastest flux emergence model and AR 12192, which has $\sim$3.6 times the flux\footnote{We perform the same calculation for AR 12192 between 2014 October 22$^\mathrm{nd}$ and 27$^\mathrm{th}$. The dataset is affected by systematic artifacts that correlate with daily spacecraft velocity variations \citep{sun2015}, so the instantaneous $\dot\Phi$ is less certain. The maximum $\Phi$, mean $\dot\Phi$, and maximum instantaneous $\dot\Phi$ are $2.18^{+0.00}_{-0.00}\times10^{23}~\mathrm{Mx}$, $8.04^{+0.26}_{-0.25}\times10^{20}\mathrm{~Mx~hr^{-1}}$, and $1.42^{+0.14}_{-0.15}\times10^{21}~\mathrm{Mx~hr^{-1}}$, respectively.}.

Our preliminary analysis suggests that AR 12673 boasts some of the highest non-potential magnetic field proxies of all ARs in Cycle 24 (X. Sun, in preparation), many of which have been used to predict flares and coronal mass ejections \citep[e.g.,][]{bobra2015}. Because major eruptions tend to occur within a day or so following significant flux emergence \citep{schrijver2009}, the emergence rate should be reconsidered as a parameter in flare prediction, as initially studied by \citet{leka2003}. AR 12673 will likely teach us more about the relations between fast flux emergence, strong non-potential field, and flaring activities.

\acknowledgments
\textit{SDO} data are courtesy of NASA and the \textit{SDO}/HMI science team.

\clearpage

\end{CJK}


\end{document}